\documentclass[pra,showpacs]{revtex4}
\usepackage{amsmath}
\usepackage{times}
\usepackage{bm}
\usepackage{graphicx}
\newcommand{\p}{{\bm r}}
\newcommand{\s}{{\bm\sigma}}
\newcommand{\tr}{\mathrm{tr}}
\newcommand{\ra}{\rangle}
\newcommand{\la}{\langle}
\newcommand{\sq}{\varrho}

\begin{document}

\title{Reduced State Uniquely Defines Groverian Measure of Original Pure State}

\author{Eylee Jung$^{1}$, Mi-Ra Hwang$^{1}$, Hungsoo Kim$^{2}$,\\
Min-Soo Kim$^{3}$, DaeKil Park$^{1,2}$, Jin-Woo Son$^{3,2}$}

\affiliation{$^1$ Department of Physics, Kyungnam University, Masan,
631-701, Korea}

\affiliation{$^2$ The Institute of Basic Science, Kyungnam University,
Masan, 631-701, Korea}

\affiliation{$^3$ Department of Mathematics, Kyungnam University, Masan,
631-701, Korea}

\email{dkpark@hep.kyungnam.ac.kr}

\author{Sayatnova Tamaryan}

\affiliation{Theory Department, Yerevan Physics Institute,
Yerevan, 375036, Armenia}

\email{sayat@mail.yerphi.am}


\begin{abstract}
Groverian and Geometric entanglement measures of the $n$-party
pure state are expressed by the $(n-1)$-party reduced state
density operator directly. This main theorem derives several
important consequences. First, if two pure n-qudit states have
reduced states of (n-1)-qudits, which are equivalent under local
unitary(LU) transformations, then they have equal Groverian and
Geometric entanglement measures. Second, both measures have an
upper bound for pure states. However, this upper bound is reached
only for two qubit systems. Third, it converts effectively the
nonlinear eigenvalue problem for three qubit Groverian measure
into linear eigenvalue equations. Some typical solutions of these
linear equations are written explicitly and the features of the
general solution are discussed in detail.
\end{abstract}
\maketitle

\section{Introduction}

Quantum theory opens up new possibilities for information
processing and communication and the entanglement of a quantum
state allows to carry out tasks, which could not be possible with
a classical system \cite{schum,Ben,benn,vedr,hor1,woot,niels,pv}.
It plays a pivotal role for exponential speedup of quantum
algorithms \cite{Vid1}, teleportation \cite{bentel} and superdense
coding \cite{dence}.

The quantum correlation is the essence of the entanglement and it
cannot be created by local operations and classical communication
(LOCC) alone. Analysis of multi-particle entanglement provides
insight into the nature of quantum correlation. However, current
situation is far from satisfaction.

Linden {\it et al.} revealed that almost every pure state of three
qubits is completely determined by its two-particle reduced
density matrices \cite{red}. In other words, we cannot get much
new information from the given pure three-qubit state if the
reduced two-qubit states are known. The case of pure  states of
any number $n$ of parties was considered in Ref.\cite{redgen} and
it was shown that the reduced states of a fraction of the parties
uniquely specify the quantum state. One may consider more general
and open questions of vital importance: how much information is
contained in any reduced $(n-1)$-qubit state? How do we use this
information to convert the {\it nonlinear eigenproblem} of
entanglement measure calculation to the {\it linear eigenproblem}?
Is there any physically relevant connection between the pure
$n$-party states which have LU-equivalent $(n-1)$-party reduced
states? Does such a connection impose an upper bound for
entanglement measure?

Groverian entanglement measure $G$ \cite{bno} gives concise
answers to all these questions. It is an entanglement measure
defined in operational terms, namely, how well a given state
serves as the input to Grover's search algorithm \cite{grov}.
Groverian measure depends on maximal success probability
$P_{\max}$ and is defined by the formula
$G(\psi)=\sqrt{1-P_{\max}}$. The maximal success probability is
the overlap of a given state with the nearest separable state. The
same overlap defines Geometric measure of entanglement introduced
earlier as an axiomatic measure\cite{Shim,barn,wei}. In this view
Groverian measure gives an operational treatment of the axiomatic
measure and is a good tool to investigate the above-mentioned
questions. In the following we will consider only the maximal
success probability and our conclusions are valid for both
Groverian and Geometric measures.

Surprisingly enough, any reduced state resulting from a partial
trace over a single qubit suffices to find $P_{\max}$ of the
original pure state. For example, the entanglement of three-qubit
pure state is completely understood from the two-qubit mixed state
reduced from the original pure state. Since bipartite systems,
regardless mixed or pure, always give a linear eigenproblem, this
fact enables us to obtain analytic expressions of Groverian
entanglement measures for pure three qubit states.

It is well-known that entanglement measures are invariant under
local unitary transformations \cite{vedr,pop,vp,hor}. However,
LU-equivalent condition is not the only one for the same Groverian
entanglement measure. In fact, if two pure states have
LU-equivalent reduced states which are obtained by taking partial trace once,
it turns out that they have
same entanglement measures. Owing to this the lower bound for
$P_{\max}$ is derived. However, it is not reachable for three and
higher qubit states and, therefore, is not precise.

In Section II we derive a formula connecting Groverian measure of
a pure state and its reduced density matrix. In Section III we
establish a lower bound for Groverian measure. In Section IV we
present analytic expressions for the maximal success probability
that reflect main features of both measures. In Section V we make
concluding remarks.

\section{Groverian measure in terms of reduced densities}

We consider a pure $n$-qudit state $|\psi\ra$. The maximum
probability of success is defined by

\begin{equation}\label{gen.pmaxpsi}
P_{\max}(\psi)=\max_{q_1q_2\ldots q_n}|\la q_1q_2\ldots
q_n|\psi\ra|^2,
\end{equation}

\noindent where $|q_k\ra$'s are pure single qudit normalized
states. Our intention is to derive a formula which connects the
maximum probability of success and $(n-1)$-qudit reduced states.
In general, reduced states are mixed states and are described by
density matrices. Hence we express the maximum probability of
success in terms of density operators right away. We will use the
notation $\rho$ for the state $|\psi\ra$ and $\sq$ for the pure
single qudit state density operators, respectively.
Eq.(\ref{gen.pmaxpsi}) takes the form

\begin{equation}\label{def.pmaxrho}
P_{\max}(\rho)=\max_{\sq_1\sq_2\ldots\sq_n}\tr
\left(\rho\,\sq_1\otimes\sq_2\otimes\cdots \otimes\sq_n\right).
\end{equation}

{\bf Theorem 1.} {\it Any $(n-1)$-qudit reduced state uniquely
determines the Groverian and Geometric measures of the original
$n$-qudit pure state.}

\smallskip

{\bf Proof.} \cite{ref} Define a single qudit state $|\chi\ra$ by
the formula

\begin{equation}\label{chi}
|\chi\ra=\la q_1q_2\ldots\widehat{q_k}\ldots q_n|\psi\ra,
\end{equation}

\noindent where \,$\widehat{}$  \,means exclusion. Obviously

\begin{equation}\label{pmax-chi}
|\la q_1q_2\ldots q_n|\psi\ra|^2=|\la
q_k|\chi\ra|^2=\tr(|\chi\ra\la\chi|\sq_k).
\end{equation}

The absolute value of the inner product $|\la q_k|\chi\ra|$ is
maximum when $q_k=|\chi\ra/\sqrt{\la\chi|\chi\ra}$ and therefore
\begin{equation}\label{ref-2}
\max_{\sq_k}\tr(|\chi\ra\la\chi|\sq_k)=\langle\chi|\chi\rangle=
\tr\left(|\chi\rangle\langle\chi|\right).
\end{equation}

Denote by $\rho(\widehat{k})$ the reduced state resulting from a
partial trace over $k$-th qudit, that is
$\rho(\widehat{k})=\tr_k\rho(\psi)$. From this definition it
follows the identity

\begin{equation}\label{chi.iden}
\tr(|\chi\ra\la\chi|)=\tr\left(\rho(\widehat{k})\sq_1\otimes
\sq_2\otimes\ldots\widehat{\sq_k}\ldots\otimes\sq_n\right).
\end{equation}

Owing to this identity Eq.(\ref{ref-2}) can be rewritten as

\begin{equation}\label{unit}
\max_{\sq^k}\tr\left(\rho\,\sq^1\otimes\sq^2\otimes\ldots\otimes
\sq^n\right)= \tr\left(\rho(\widehat{k})\sq_1\otimes
\sq_2\otimes\ldots\widehat{\sq_k}\ldots\otimes\sq_n\right).
\end{equation}

Both sides of the Eq.(\ref{unit}) must have the same maximum and
this is the proof of the theorem.

Since the r.h.s. of Eq.(\ref{unit}) contains the reduced density
operator $\tr_k\,\rho=\rho(\widehat{k})$ which is generally mixed
state, the next maximization is nontrivial.

Eq.(\ref{unit}) does not mean that a pure state and its once
reduced state have equal Groverian measures. One can not maximize
the mixed state density matrix over product states to find the
entanglement measure because the resulting measure is not an
entanglement monotone\cite{bno,wei,shap}.

Eq.(\ref{unit}) connects directly the maximum probability of
success with the reduced density operator

\begin{equation}\label{reduced}
P_{\max}(\rho)=\max_{\sq_1\sq_2\ldots\widehat{\sq_k}\ldots\sq_n}
\tr\left(\rho(\widehat{k})\sq_1\otimes
\sq_2\otimes\ldots\widehat{\sq_k}\ldots\otimes\sq_n\right).
\end{equation}

In fact, Theorem\,1 is true for any entanglement measure
\cite{gyot}. Consider  an (n-1)-qudit reduced density matrix that
can be purified by a single qudit reference system. Let
$|\psi^\prime\ra$ be any joint pure state. All other purifications
can be obtained from the state $|\psi^\prime\ra$ by
LU-transformations $U\otimes\openone^{\otimes(n-1)}$ where $U$ is
a local unitary matrix acting on single qudit and $\openone$ is a
unit matrix. Since any entanglement measure must be invariant
under LU-transformations, it must be the same for all
purifications independently of $U$. Hence the reduced density
matrix $\rho$ determines any entanglement measure on the initial
pure state.

However, there is a crucial difference. In the case of Groverian
measure the proof expresses entanglement measure by the reduced
density matrix directly. As will be explained in Section IV,
Eq.(\ref{reduced}) is a simple and effective tool for calculating
three-qubit entanglement measure. No such formula is known for
other measures and general proof for other measures has limited
practical significance.

{\bf Theorem 2.} {\it If two pure $n$-qudit states have LU
equivalent $(n-1)$-qudit reduced states, then they have equal
Groverian and Geometric entanglement measures.}

\smallskip

{\bf Proof.} Assume that the density matrices of pure states are
$\rho$ and $\rho^\prime$ and corresponding maximum probabilities
of success are $P_{\max}$ and $P_{\max}^\prime$. Suppose the local
unitary transformation $U^1\otimes U^2\otimes\cdots\otimes
U^{n-1}$ maps
$\rho^\prime(\widehat{k^\prime})=\tr_{k^\prime}\rho^\prime$ to
$\rho(\widehat{k})=\tr_k\rho$ as following:

\begin{equation}\label{map}
\rho(\widehat{k})=\left(U^1\otimes U^2\otimes\cdots\otimes
U^{n-1}\right) \rho^\prime(\widehat{k^\prime})\left(U^1\otimes
U^2\otimes\cdots\otimes U^{n-1}\right)^{+},
\end{equation}

\noindent where superscript $+$ means hermitian conjugate. The
trace with any complete product
$\sq^1\otimes\sq^2\otimes\cdots\otimes\sq^{n-1}$ state gives

\begin{equation}\label{map.con}
\tr\left(\rho(\widehat{k})\sq^1\otimes\sq^2\otimes\cdots\otimes\sq^{n-1}
\right)= \tr\left(\rho^\prime(\widehat{k^\prime})
\sq^{\prime1}\otimes\sq^{\prime2}\otimes\cdots\otimes\sq^{\prime
n-1} \right),
\end{equation}

\noindent where $\sq^{\prime k}=U^{k+}\sq^kU^k$ are single qubit
pure states too. Let's choose the product state that maximizes the
l.h.s. According to Eq.(\ref{reduced}) l.h.s is $P_{\max}$ and
therefore $P_{\max}\leq P_{\max}^\prime$. Similarly
$P_{\max}^\prime\leq P_{\max}$, therefore $P_{\max}=
P_{\max}^\prime$.

\section{Lower bound for multi-qubit systems}

Theorem\,1 sets a clear lower bound  for the maximum probability
of success.

Below $A$ is an arbitrary $2\times2$ hermitian matrix, $\p$ is a
unit real three-dimensional vector and components of the vector
$\s$ are Pauli matrices. The trace of the product of matrices $A$
and $\p\cdot\s$ can be presented as a scalar product of vectors
$\p$ and $\tr(A\s)$. The scalar product of two real vectors with
the constant modules is maximal when vectors are parallel.
Consequently, we have

\begin{equation}\label{start}
\max_{r^2=1}\,\tr\left(A\, \p\cdot\s\right)=|\tr(A\s)|=\sqrt{(\tr
A)^2-4\det A}
\end{equation}

\noindent and the positive root of radicals is understood.

An arbitrary density matrix $\sq$ for a pure state qubit may be
written as $\sq=1/2\,(\openone+\p\cdot\s)$, where  and $\p$ is a
unit real vector. Then Eq.(\ref{start}) can be rewritten as

\begin{equation}\label{maximum}
\max_{\sq}\,\tr\left(A\,\sq\right)= \frac{1}{2}\left(\tr
A+\sqrt{(\tr A)^2-4\det A}\right).
\end{equation}

From Eq.(\ref{maximum}) it follows that

\begin{equation}\label{start.ineq}
\max_{\sq}\,\tr\left(A\,\sq\right)\geq\frac{1}{2}\left(\tr
A\right).
\end{equation}

We define $2\times2$ matrix $M_{n-1}$ by formula

\begin{equation}\label{mn}
M_{n-1}=\tr_{1,2,...,n-2}\left(\rho(\widehat{n})\sq^1\otimes\sq^2,
\otimes\cdots\sq^{n-2}\otimes\openone\right).
\end{equation}

\noindent where trace is taken over (1,2,...,n-2)-qubits.
Eq.(\ref{reduced}) takes the form

\begin{equation}\label{p-mn}
P_{\max}=\max_{\sq^1\sq^2\cdots\sq^{n-1}}\tr(M_{n-1}\sq^{n-1}),
\end{equation}

\noindent where $\tr$ means trace over (n-1)-qubit.
Eq.(\ref{start.ineq}) gives

\begin{equation}\label{ineq-1}
P_{\max}\geq\frac{1}{2}\max_{\sq^1\sq^2\cdots\sq^{n-2}}\tr
M_{n-1}=\frac{1}{2}\max_{\sq^1\sq^2\cdots\sq^{n-2}}
\tr\left(\rho(\widehat{n})\sq^1\otimes\sq^2
\otimes\cdots\sq^{n-2}\otimes\openone\right),
\end{equation}

\noindent where $\tr$ in rhs of Eq.(\ref{ineq-1}) means trace over
all qubits.Thus inequality (\ref{start.ineq}) suggests a simple
prescription: replace a pure qubit density matrix by unit matrix
and add a multiplier $1/2$ instead. We use this prescription $n-1$
times, eliminate all single qubit density operators step by step
from Eq.(\ref{reduced}) and obtain

\begin{equation}\label{basic.ineq}
 P_{\max}\geq\frac{1}{2^{n-1}}.
\end{equation}

\noindent Note that this lower bound is valid only for pure
states. The question at issue is whether it is a precise limit or
not. And if it is indeed the case, then what are the pure states
which have the lower bound of $P_{\max}$ ? We will prove that this
lower bound is reached only for bipartite states.

Denote by $\rho^{k_1k_2\cdots k_m}$ the reduced density operator
of qubits $k_1k_2\cdots k_m,\quad1\leq m\leq n-1$. Eq.
(\ref{unit}) and (\ref{start.ineq}) together yield

\begin{equation}\label{pure-red}
P_{\max}(\rho)\geq\frac{1}{2^{n-m-1}}P_{\max}(\rho^{k_1k_2\cdots
k_m}).
\end{equation}

Note, $P_{\max}(\rho^{k_1k_2\cdots k_m})$ does not define any
entanglement measure as $\rho^{k_1k_2\cdots k_m}$'s are mixed
states. It is the maximal overlap of the mixed state with any
product state and we use it as intermediate mathematical quantity.

\smallskip

{\bf Lemma 2.} {\it If a pure state has limiting Geometric /
Groverian entanglement $P_{\max}=1/2^{n-1}$, then all its reduced
states are completely mixed states.}

\smallskip

{\bf Proof.} Eq.(\ref{pure-red}) for $m=1$ and Eq.(\ref{maximum})
impose

\begin{equation}\label{fin.ineq}
P_{\max}\geq\frac{1}{2^{n-1}}\left(1+\sqrt{1-4\det \rho^k}\right).
\end{equation}

The maximal probability of success reaches the minimal value if
the square root vanishes. Consequently, density matrices $\rho^k$
must be multiple of a unit matrix $\rho^k=\openone/2$ and thus all
one-qubit reduced states are completely mixed. Then two qubit
density matrices $\rho^{k_1k_2}$ must have the form

\begin{equation}\label{d2.cond}
\rho^{k_1k_2}=\frac{1}{4}\left(\openone\otimes\openone+
g_{\alpha\beta}\,\s^\alpha\otimes\s^\beta\right).
\end{equation}

\noindent where
$g_{\alpha\beta}=\tr(\rho^{k_1k_2}\s^\alpha\otimes\s^\beta)$ is a
$3\times3$ matrix with real entries. Hereafter summation for
repeated three dimensional vector indices
$(\alpha,\beta,\gamma\cdots=1,2,3)$ is understood unless otherwise
stated. To reach the lower bound we must have equality instead of
inequality in (\ref{pure-red}) and this condition imposes
$P_{\max}(\rho^{k_1k_2})=1/4$ resulting in $g_{\alpha\beta}=0$.
Hence $\rho^{k_1k_2}=(1/4)\openone\otimes\openone$ and thus all
two-qubit reduced states are completely mixed. One can continue
this chain of derivations by induction. Indeed, suppose all
$m$-qubit states $(m<n)$ are completely mixed. Then $(m+1)$-qubit
density matrices $\rho^{k_1k_2\cdots k_{m+1}}$ must have the form

\begin{equation}\label{induc}
\rho^{k_1k_2\cdots
k_{m+1}}=\frac{1}{2^{m+1}}\left(\openone^{\otimes
m+1}+g_{\alpha_1\alpha_2\cdots\alpha_{m+1}}
\sigma^{\alpha_1}\otimes\sigma^{\alpha_2}\otimes\cdots
\otimes\sigma^{\alpha_{m+1}}\right),
\end{equation}

\noindent where

\begin{equation}\label{gggg}
g_{\alpha_1\alpha_2\cdots\alpha_{m+1}}=\tr\left(
\rho^{k_1k_2\cdots k_{m+1}}
\sigma^{\alpha_1}\otimes\sigma^{\alpha_2}\otimes\cdots
\otimes\sigma^{\alpha_{m+1}}\right).
\end{equation}

\noindent From Eq.(\ref{pure-red}) it follows that
$P_{\max}(\psi)$ takes its minimal value if
$P_{\max}(\rho^{k_1k_2\cdots k_m})=1/2^m$.  Eq.(\ref{induc}) is
consistent with this condition if and only if the maximization of
the term  of $g_{\alpha_1\alpha_2\cdots\alpha_{m+1}}
\sigma^{\alpha_1}\otimes\sigma^{\alpha_2}\otimes\cdots
\sigma^{\alpha_{m+1}}$ yields zero. Then
$g_{\alpha_1\alpha_2\cdots\alpha_{m+1}}=0$ and therefore

\begin{equation}\label{induc-fin}
\rho^{k_1k_2\cdots k_{m+1}}=\frac{1}{2^{m+1}}\openone^{\otimes
m+1}.
\end{equation}

\noindent Thus if all $m$-qubit reduced states are completely
mixed then all $(m+1)$-qubit reduced states are also completely
mixed. On the other hand all one-qubit reduced state are
completely mixed. By induction all reduced states are completely
mixed. The induction stops at pure states. In contrast to mixed
states, the maximization of the term
$g_{\alpha_1\alpha_2\cdots\alpha_n}
\sigma^{\alpha_1}\otimes\sigma^{\alpha_2}\otimes\cdots
\sigma^{\alpha_n}$ must yield unity for pure states as requires
Eq.(\ref{unit}).

 Lemma is proved.

\smallskip

{\bf Theorem 3.} {\it None of multi-qubit pure states except
two-qubit maximally entangled states satisfies the condition
$P_{\max}=1/2^{n-1}$}.

\smallskip

{\bf Proof.} When $n=2$, it is well-known that the EPR states and
their LU-equivalent class reach the lower bound, {\it i.e.}
$P_{max} = 1/2$. Now we would like to show that there is no pure
state with limiting Groverian measure for $n=3$. Lemma 2 requires
that the density matrix with limiting Groverian measure should be
in the form

\begin{equation}
\label{three1}
\rho = \frac{1}{8}
\left(\openone^{\otimes 3} + g_{\alpha \beta \gamma}
         \sigma^{\alpha} \otimes \sigma^{\beta} \otimes \sigma^{\gamma}
                                                 \right).
\end{equation}

\noindent Since $\rho$ is a pure state density matrix, it must
satisfy $\rho^2 = \rho$. This condition leads several constraints,
one of which is
\begin{equation}\label{three2}
-i g_{\alpha \beta \gamma} g_{\delta \kappa
\lambda} \epsilon_{\alpha \delta \delta'} \epsilon_{\beta \kappa
\kappa'} \epsilon_{\gamma \lambda \lambda'}
\sigma^{\delta'}\otimes\sigma^{\kappa'}\otimes\sigma^{\lambda'}
 = 6 g_{\alpha \beta \gamma} \sigma^{\alpha} \otimes
\sigma^{\beta} \otimes \sigma^{\gamma}
\end{equation}

\noindent where $\epsilon_{\alpha \beta \gamma}$ is an
antisymmetric tensor. Since this constraint cannot be satisfied
for real $g_{\alpha \beta \gamma}$, there is no pure state which
has limiting Groverian measure at $n=3$.

Now we will show that there is no pure state for $n \geq 4$ too.
Suppose there is $n$-qubit state $|\psi\ra$ such that all its
reduced states are completely mixed. Choose a normalized basis of
product vectors $|i_1i_2\cdots i_n\ra$ where the labels within ket
refer to qubits $1,2,\cdots n$ in that order. The vector
$|\psi\ra$ can be written as a linear combination

\begin{equation}\label{psi}
|\psi\ra=\sum_{i_1i_2\cdots i_n}C_{i_1i_2\cdots i_n}|i_1i_2\cdots
i_n\ra
\end{equation}

\noindent of vectors in the set. All reduced states of the state
$|\psi\ra$  are completely mixed if and only if

\begin{equation}\label{cond1}
\sum_{i_kj_k}\delta_{i_kj_k}C_{i_1i_2\cdots i_n}C_{j_1j_2\cdots
j_n}^\ast= \frac{1}{2^{n-1}}\delta_{i_1j_1}\delta_{i_2j_2}\cdots
\widehat{\delta_{i_kj_k}}\cdots\delta_{i_nj_n},\quad k=1,2,\cdots
n.
\end{equation}

\noindent Note that normalization condition follows from above
equation. Define $n-1$ index coefficients

\begin{equation}\label{down}
D_{i_1i_2\cdots i_{n-1}}=\sqrt{2}C_{i_1i_2\cdots i_{n-1}0}.
\end{equation}

Setting $i_n=j_n=0$ in Eq.(\ref{cond1}) we get

\begin{equation}\label{cond2}
\sum_{i_kj_k}\delta_{i_kj_k}D_{i_1i_2\cdots
i_{n-1}}D_{j_1j_2\cdots j_{n-1}}^*=
\frac{1}{2^{n-2}}\delta_{i_1j_1}\delta_{i_2j_2}\cdots
\widehat{\delta_{i_kj_k}}\cdots\delta_{i_{n-1}j_{n-1}},\;
k=1,2,\cdots n-1.
\end{equation}

Hence the $(n-1)$-qubit state

\begin{equation}\label{psilow}
|\phi\ra=\sum_{i_1i_2\cdots i_{n-1}}D_{i_1i_2\cdots
i_{n-1}}|i_1i_2\cdots i_{n-1}\ra
\end{equation}

\noindent exists and all its reduced states are completely mixed.
The contraposition of it is that if there is no pure state which
has limiting Groverian measure at $n=3$, it is also true for $n
\geq 4$. Theorem\,3 is proved.

\bigskip

Thus, the lower bound of inequality (\ref{basic.ineq}) is
unreachable for $n \geq 3$. This seems to mean that
Eq.(\ref{basic.ineq}) is not a precise limit.

\section{Analytic expressions for maximum probability of success}

The maximization of the pure three qubit states over product
states generally reduces to {\it nonlinear} eigenvalue equations
\cite{wei}. However, Eq.(\ref{reduced}) converts it effectively
into {\it linear} eigenvalue equations. Thus, one can compute the
entanglement measures for wide range of three qubit states
analytically. As an illustration consider one parametric W-type
\cite{Chir} three qubit state

\begin{equation}\label{example}
|\psi\ra = \frac{1}{\sqrt{1 + \kappa^2 + \kappa^4}} (|100\ra +
\kappa |010\ra + \kappa^2 |001\ra),
\end{equation}

\noindent where $\kappa$ is a free positive parameter.
The calculation method is elaborated in Ref.\cite{tri} and here we
present only final results. In three different ranges of
definition the maximal success probability is differently
expressed.
In the first case $P_{\max}$ is the square of the first
coefficient provided it is greater than $1/2$:

\begin{equation}\label{exam-1}
P_{\max}=\frac{1}{1+\kappa^2+\kappa^4},\quad
0<\kappa<\left(\frac{\sqrt{5}-1}{2}\right)^{1/2}.
\end{equation}

In the second case $P_{\max}$ is the square of the diameter of the
circumcircle of the acute triangle formed by three coefficients:

\begin{equation}\label{exam-r}
P_{\max}=\frac{4\kappa^6}{(1+\kappa^2+\kappa^4)^2(3\kappa^2-1-\kappa^4)},\quad
\left(\frac{\sqrt{5}-1}{2}\right)^{1/2}\leq \kappa\leq
\left(\frac{\sqrt{5}+1}{2}\right)^{1/2}.
\end{equation}

In the third case $P_{\max}$ is the square of the third
coefficient provided it is greater than $1/2$:

\begin{equation}\label{exam-k}
P_{\max}=\frac{\kappa^4}{1+\kappa^2+\kappa^4},\quad
\kappa>\left(\frac{\sqrt{5}+1}{2}\right)^{1/2}.
\end{equation}

It is also possible to compute $P_{\max}$ for Eq.(\ref{example})
numerically\cite{Shim-grov}. For numerical calculation we consider
$k^{th}$ qubit as $|q_k\rangle = \cos \theta_k |0\rangle + e^{i
\varphi_k} \sin \theta_k |1\rangle$ with $k=1, 2, 3$. Since the
coefficients of $|\psi\rangle$ are all real, we can put $\varphi_k
= 0$ for all $k$ and express $P_{\max}$ in a form

\begin{equation}
\label{numer1} P_{max} = \max_{\theta_1,\theta_2,\theta_3} |\la
q_1|\la q_2|\la q_3|\psi \ra |^2.
\end{equation}

Thus numerical maximization over $\theta_1$, $\theta_2$ and
$\theta_3$ directly yields $P_{\max}$. As shown in Fig. 1(a) the
numerical result (black dots) perfectly coincides with the
analytic results (solid lines) expresses in Eq.(\ref{exam-1}),
(\ref{exam-r}) and (\ref{exam-k}).

\begin{figure}
\includegraphics[width=8cm]{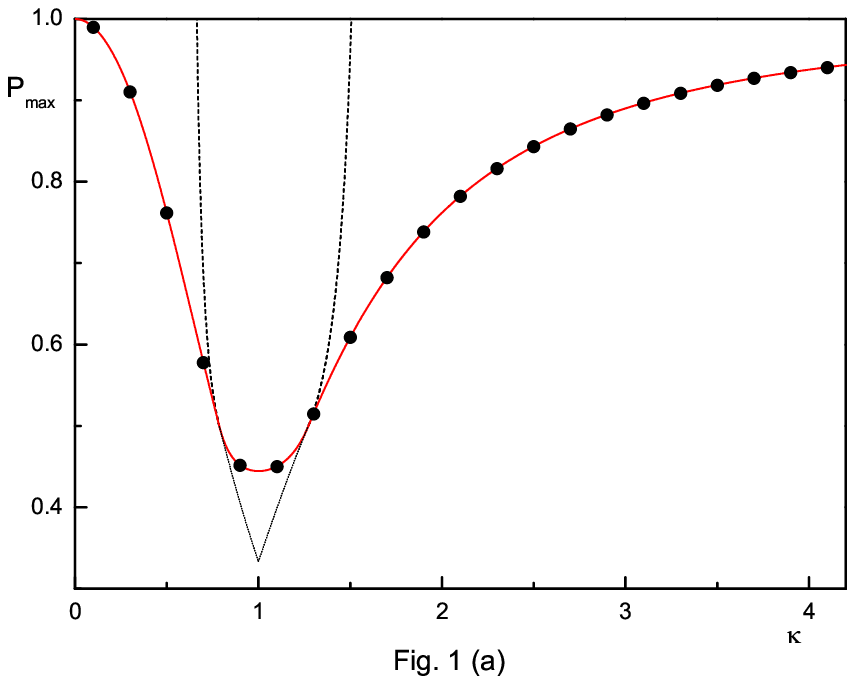}
\includegraphics[width=8cm]{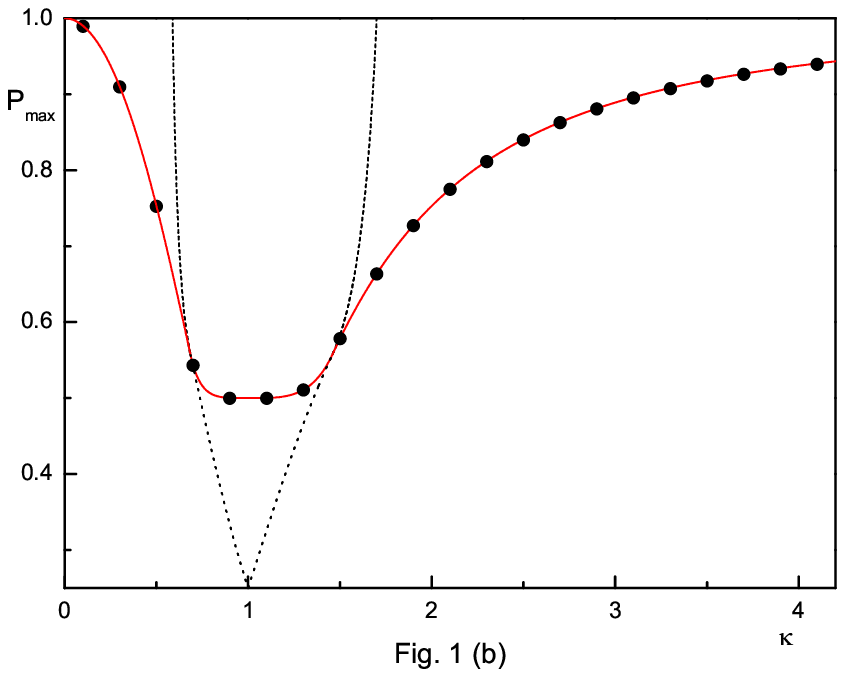}
\caption[fig1]{(Color
online) $P_{max}$ for Eq.(\ref{example}) (Fig. 1 a) and
Eq.(\ref{example-2}) (Fig. 1 b). The solid lines represent the
analytical results of $P_{max}$ and the black dots are the
numerical results. This figures strongly support that our
analytical results are perfect correct.}
\end{figure}

Let us consider another one parametric state

\begin{equation}\label{example-2}
|\psi\ra = \frac{1}{\sqrt{1 + \kappa^2 + \kappa^4 + \kappa^6}}
\left(|100\ra + \kappa |010\ra + \kappa^2 |001\ra + \kappa^3
|111\ra\right).
\end{equation}

Again there are three cases. If four coefficients form a cyclic
quadrilateral, then $P_{\max}=4R^2$, where $R$ is the circumradius
of the quadrangle. Otherwise $P_{\max}$ is the square of the
largest coefficient. In the first case $P_{\max}$ is the square of
first coefficient:

\begin{eqnarray}\label{exam2-1}
 & & P_{\max}=\frac{1}{1 + \kappa^2 + \kappa^4 +
 \kappa^6},\\\nonumber
 & & \kappa<\frac{1}{3}\left(\sqrt[3]{18\sqrt{57}+134}-
\sqrt[3]{18\sqrt{57}-134}-1\right)^{1/2}\approx0.685.
\end{eqnarray}

In the second case $P_{\max}$ is the square of the circumcircle of
the cyclic quadrangle formed by four coefficients:

\begin{eqnarray}\label{exam2-r}
 & &
P_{\max}=\frac{8\kappa^6}
{-1+2\kappa^2+\kappa^4+8\kappa^6+\kappa^8+2\kappa^{10}
-\kappa^{12}},\\\nonumber
 & & \frac{1}{3}\left(\sqrt[3]{18\sqrt{57}+134}-
\sqrt[3]{18\sqrt{57}-134}-1\right)^{1/2}\leq \kappa\leq
\frac{1}{\sqrt{3}}\left(\sqrt[3]{46+6\sqrt{57}}+
\sqrt[3]{46-6\sqrt{57}}+1\right)^{1/2}.
\end{eqnarray}

In the third case $P_{\max}$ is the square of the last
coefficient:

\begin{eqnarray}\label{exam2-4}
 & & P_{\max}=\frac{\kappa^6}{1 + \kappa^2 + \kappa^4 +
 \kappa^6},\\\nonumber
 & & \kappa>\frac{1}{\sqrt{3}}\left(\sqrt[3]{46+6\sqrt{57}}+
\sqrt[3]{46-6\sqrt{57}}+1\right)^{1/2}\approx1.46.
\end{eqnarray}

The function $P_{\max}(k)$ and numerical results are shown in Fig.
1(b). Both figures strongly show that our analytical expressions
of $P_{\max}$ perfectly coincide with the numerical result.

\section{Conclusions}

Eq.(\ref{reduced}) allows to calculate the maximal success
probability for three qubit states which are expressed as linear
combinations of four given orthogonal product states
\cite{shared}. The answer is more complicated than a simple
formula, but each final expression of the measure has its own
meaningful interpretation. Namely, $P_{\max}$ can take the
following values(up to numerical coefficients):

\begin{itemize}

\item the square of the circumradius of the cyclic polygon formed
by coefficients of the state function,

\item the square of the circumradius of the crossed figure formed
by coefficients of the state function,

\item the largest coefficient.

\end{itemize}

Each expression has its own range of definition where they are
applicable. Although the above picture seems simple, the
separation of the applicable domains is highly nontrivial task. To
make clear which of expressions should be applied for a given
state we refer to \cite{shared}. All our results on Groverian
measure of three qubit pure states are summarized in \cite{3qub}.

Eq.(\ref{reduced}) gives nonlinear eigenvalue problem for four and
higher qubit states and it is natural to ask whether there is an
extension of Eq.(\ref{reduced}) that allows to find analytic
results for four, five, or general n-qubits. Although we have no
distinct results here, but we have obtained some insight from the
analysis of the information contained in one and two qubit reduced
states. Probably, it is possible to express the maximal success
probability in terms of one and two qubit reduced states in case
of four qubit pure states. Such formula, if it can be derived, will
give linear equations for four qubit pure states. However,
situation is opposite in the case of five qubit states. The method
does now allow to convert the task to the linear eigenvalue
problem and more powerful tools are needed to calculate maximal
success probability of general n-qubit states.

\bigskip

\begin{acknowledgments} We thank Levon Tamaryan for valuable
discussions.

This work was supported by the Kyungnam University Research Fund,
2007.
\end{acknowledgments}

\end{document}